\documentclass[english,aps,10pt]{article}

\usepackage[T1]{fontenc}
\usepackage[latin1]{inputenc}
\usepackage{graphicx}
\usepackage{amssymb}
\usepackage{sublabel}
\usepackage{epsfig}
\usepackage{simplemargins}

\newcommand{\comment}[1]{}

\setallmargins{1in}

\newcommand{\captionfonts}{\scriptsize}

\makeatletter  
\long\def\@makecaption#1#2{%
  \vskip\abovecaptionskip
  \sbox\@tempboxa{{\captionfonts #1: #2}}%
  \ifdim \wd\@tempboxa >\hsize
    {\captionfonts #1: #2\par}
  \else
    \hbox to\hsize{\hfil\box\@tempboxa\hfil}%
  \fi
  \vskip\belowcaptionskip}
\makeatother   

\makeatletter

\usepackage{babel}
\makeatother
\begin{document}

\title{Uninformative memories will prevail: the storage of correlated representations and its consequences}
\author{Emilio Kropff and Alessandro Treves}
\comment{\email{kropff@sissa.it}
\homepage{http://people.sissa.it/~kropff}
\affiliation{SISSA, Cognitive Neuroscience \\
 via Beirut 2-4 \\
 34014 Trieste, Italy}
\author{Alessandro Treves}
}
 \maketitle
\begin{abstract}
Autoassociative networks were proposed in the 80's as simplified models of memory function in the brain, using recurrent connectivity with hebbian plasticity to store patterns of neural activity that can be later recalled. This type of computation has been suggested to take place in the CA3 region of the hippocampus and at several levels in the cortex. One of the weaknesses of these models is their apparent inability to store correlated patterns of activity. We show, however, that a small and biologically plausible modification in the `learning rule' (associating to each neuron a plasticity threshold that reflects its \textit{popularity}) enables the network to handle correlations. We study the stability properties of the resulting memories (in terms of their resistance to the damage of neurons or synapses), finding a novel property of autoassociative networks: not all memories are equally \textit{robust}, and the most \textit{informative} are also the most sensitive to damage. We relate these results to category-specific effects in semantic memory patients, where concepts related to `non-living things' are usually more resistant to brain damage than those related to `living things', a phenomenon suspected to be rooted in the correlation between representations of concepts in the cortex.

  Total number of words: 9809
  
 Total number of characters (including spaces): 70703
\end{abstract}


\section{Introduction}
Autoassociative memory networks can store patterns of neural activity by modifying the synaptic weights 
that interconnect neurons \cite{Hopfield, Amit_1989}, following the simple rule first stated by Donald 
O. Hebb: \textit{neurons that fire together wire together} \cite{Hebb}. Once a pattern of activity is 
stored, it becomes an attractor of the dynamics of the system. Evidence of attractor behavior has been reported in 
the rat hippocampus \textit{in vivo} \cite{OKeefe_2005}. Such memory mechanisms 
have been proposed to be present throughout the cortex, where hebbian plasticity plays a major role. 

The theoretical and computational literature studying variations of the original Hopfield model 
\cite{Hopfield} is profuse. Advantages toward optimality or biological plausibility have been 
demonstrated by varying the learning rule, the neuron model, the architecture or connectivity scheme
and the statistics of the input data. The resulting changes in the behavior of the network, however, are 
often quantitative rather than qualitative. Attractor networks are robust systems that depend only 
weakly on details. Any optimized attractor network, in fact, appears to be able to retrieve a total 
amount of information that is never more than a fraction of a bit per synaptic variable. This limit,
consistent with insight obtained with the Gardner approach \cite{Gardner_1988} but never fully proven, 
implies that the `storage capacity' of any associative memory network is constrained by the number 
of independently modifiable synapses it is endowed with. A suboptimal organization can easily underutilize
such capacity, but no clever arrangement can do better than that. Crossing the capacity limit induces a 
`phase transition' into total amnesia, destroying the attractor dynamics that would lead to memory states.

Subtler memory deficits than an overall collapse have been reported in the neuropsychological 
literature, such as category specific effects in the semantic memory system. Patients with partial 
damage in the cortical networks sustaining semantic memory are found to lose preferentially 
some concepts rather than others (typically \textit{animals} rather than \textit{tools} or \textit{living} 
rather than \textit{non-living} things). Initially, research on these effects produced two major antagonistic accounts: the sensory-functional theory \cite{Warrington_Shallice_1984, Warrington_McCarthy_1987} and the domain specific theory \cite{AlfonsoCaramazza01011998}. Roughly, they hypothesize that different categories of concepts are localized within partially different (the former) or completely different (the latter) cortical networks. Damage to particular areas would then produce a deficit in the corresponding category of concepts. Attempts to validate some predictions of these theories have not been successful, and an alternative view has emerged in the last few years that, although formulated in various ways, basically hypothesizes that the crucial factor to understand category specific effects is the correlation among items of semantic information, presumed to be stored in one extended and only weakly heterogeneous network \cite{Tyler_2002,Tyler_2000, Sartori_2004, McRae_1997}. According to this view, random damage to the network would produce selective impairments not because one category is more localized within the damaged area than the other, but rather because differences in the structure of correlations make some categories more vulnerable to damage than others. This explanation has been formulated in a qualitative rather than quantitative formulation. The object of the present study is to fill this gap with a theory that produces systematic quantitative predictions applicable, in principle, to these and other memory networks storing correlated information. We focus on mathematical models that allow to assess the hypothesis in its `pure' form, without discussing further other accounts of category specific deficits, found in the literature, which may of course offer complementary elements to an integrated explanation of empirical results.

Most models of attractor networks consider patterns that, for the sake of the analysis, are generated by a simple random process, uncorrelated with each other. Some exceptions appeared during the 80's, when interest grew around the storage 
of patterns derived from hierarchical trees \cite{Parga_1986, Gutfreund_1988}. In particular, Virasoro \cite{Virasoro_1988} relates the behavior of networks of general architecture to \textit{prosopagnosia}, an 
impairment in certain patients to identify individual stimuli (e.g., faces) but not to categorize them. Interestingly, his model indicates that prosopagnosia is not prevalent in 
networks endowed with Hebbian-plasticity. Other developments have described perceptron-like or other 
local rules to store generally correlated patterns \cite{Gardner_1989, Diederich_1987, Srivastava_2004} or patterns 
with specifically spatial correlation \cite{Monasson_1992}. More recently, Tsodyks and collaborators \cite{Tsodyks_2006} have studied a Hopfield memory in which a sequence of morphs between two 
uncorrelated patterns is stored. In their work, the use of a saliency function favouring unexpected 
over expected patterns, during learning, can result in the formation of a continuous one-dimensional 
attractor that spans the space between two original memories. Such fusion of basins of attraction is 
an interesting phenomenon that we leave for a later extension of this work.
In this report, we assume that the elements stored in semantic memory are discrete by construction.

In summary, we aim to show here how a modified version of the standard `Hebbian' plasticity rule enables an autoassociative network to store and retrieve correlated memories, and how a side effect of the need to use this modified learning rule is the emergence of substantial variability in the resistance of individual memories to damage, which, as we discuss, could explain the prevailing trends of category specific memory impairments observed in patients.

\subsection{Attractor networks}

Attractor networks are thought to sustain memory at several levels in the cortex and hippocampus, by virtue of recurrent connections endowed with hebbian plasticity. Models consider input information to the system to be organized into patterns of activity, which the network has to `remember'. We represent these patterns by means of the variables $\xi_i^{\mu}$, which stand for the activity of neuron $i$ in the network when pattern $\mu$ is being fed as an input. The weight of each recurrent synapse is modified following the coactivation of the pre and post synaptic neurons. In the simplest model, neurons that were strongly activated by the presentation of pattern $\mu$ reinforce their mutual connections, as a result of which if only a group of them is active at some time in the future, the others also tend to be activated. In other words, the presentation of a `cue' causes the retrieval of the whole memory, which is a stable firing state of the network, also called an attractor of its dynamics. 

While some studies model the learning process itself, in which patterns are presented as inputs and synapses modified, others assume that learning has already occurred, so that stable or ideal weights have been reached, and analyze the resulting performance of the network. The present work belongs to this second group.

If several patterns are memorized in the same network, the modifications introduced by each of them may be added linearly to the weight of synapses. When the total number of stored patterns $p$ is large enough, such that neurons and synapses are shared by many different patterns, any attempt to retrieve a memorized pattern could suffer from `interference', understood as the summed effect of the other memorized patterns on the relevant synapses. Theoretical studies have shown that in a network storing random patterns, the strength of this interference depends on the parameter $\alpha\equiv p/C$, where $C$ is the mean number of afferent connection weights to each neuron. If the memory load is small and negligible, $\alpha\sim 0$, memories are retrieved optimally, or, in other words, the original patterns of activity are themselves stable attractors of the system. When $\alpha$ is not negligible but still smaller than a critical value $\alpha_c$ (the \textit{storage capacity} of the network), patterns can be retrieved but not optimally. If a partial cue of pattern $\mu$ is presented to the network, its activity evolves to a stable attractor state presenting a high but not full overlap with the original pattern. The interference is not destructive, but displaces the attractors slightly out of their original positions. As $\alpha$ increases, approaching $\alpha_c$, this effect is stronger: the overlap between the attractor and the original pattern is progressively lower, and the capability to complete partial cues is diminished. In the limit of $\alpha = \alpha_c$, attractors are stable but the network does not evolve towards them; retrieval occurs only when the cue is already the full attractor. Finally, when $\alpha>\alpha_c$, the attractors become unstable and the stored memories are no longer retrievable.

\subsection{The model}

We consider a network with $N$ neurons and $C< N$ afferent synaptic connections per neuron. The network stores $p$ patterns, and the parameter $\alpha = p/C$ measures its memory load. As for classical analyses \cite{Amit_1989}, we take the `thermodynamic' limit ($p\rightarrow \infty$, $C\rightarrow \infty$, $N\rightarrow \infty$, $\alpha$ constant, $C/N$ constant) in which the equilibrium properties of the network depend on $\alpha$ rather than separately on $N, C$ and $p$.

The activity of neuron $i$ is described by the variable $\sigma_i$, with $i=1...N$. Each of the $p$ patterns is a particular state of activation of the network. The activity of neuron $i$ in pattern $\mu$ is described by $\xi_i^{\mu}$, with $\mu=1...p$. The perfect retrieval of pattern $\mu$ is thus characterized by $\sigma_i=\xi_i^{\mu}$ for all $i$. For the sake of simplicity, we will assume binary patterns, where $\xi_i^{\mu}=0$ if the neuron is silent and $\xi_i^{\mu}=1$ if the neuron fires. Consistently, the activity states of neurons will be limited by $0\leq \sigma_i\leq 1$. Extensions of this work to e.g. threshold-linear units \cite{Treves_1990} or to Potts units \cite{potts-2005} are left for further analyses, though, as usual with attractor networks, there is no reason to expect large differences in the qualitative behavior of the system. 

We assume that a fraction $a$ of the neurons is activated in each pattern, $a=\Sigma_i{\xi_i^{\mu}}/N$ for $\mu=1 \dots p$. This $sparseness$ parameter is critical in determining the storage capacity of any associative memory network \cite{Treves_1991}.

Each neuron receives $C$ synaptic inputs. To describe the architecture of connections we use a random matrix with elements $c_{ij}=1$ if a synaptic connection between post-synaptic neuron $i$ and pre-synaptic neuron $j$ exists and $c_{ij}=0$ otherwise, with $c_{ii}=0$ for all $i$, a requirement for most attractor network models to function. In addition, synapses have associated weights $J_{ij}$. 

The influence of the network activity on a given neuron $i$ is represented by the field
\begin{equation}
h_i=\sum_{j=1}^N c_{ij}J_{ij}\sigma_j \label{eq:field}
\end{equation}
which enters a sigmoidal activation function when updating the activity of the neuron
\begin{equation}
\sigma_i=\left\{1+\exp \beta \left(U-h_i\right)\right\}^{-1}\label{eq:activation}
\end{equation}
where $\beta$ is an inverse temperature parameter and $U$ is a threshold parameter, which must be kept of order $1$ (given the appropriate scaling of the weigths that we will adopt) in order to have a storage capacity close to optimal \cite{Buhmann_1989, Tsodyks_1988}. If $U\ll 1$ all the neurons tend to activate, somewhat similarly to what happens during an epileptic seizure. If, on the other extreme, $U\gg 1$, all neurons tend to be silent. In both extreme situations the effect of $U$ on the network is much stronger than that of the attractors. When $U$ is of order $1$, on the contrary, the attractors dominate the dynamics of the network, keeping the total activity of the network near the sparseness $a$ even for transient states, independently of small variations of $U$.

The learning rule that defines the weights $J_{ij}$ in classical models reflects the Hebbian principle: every pattern in which both neurons $i$ and $j$ are active contributes positively to $J_{ij}$. In addition, in order to optimize storage, 
the rule may include some prior information about pattern statistics. In a one-shot learning paradigm, with uncorrelated patterns, the optimal rule uses the sparseness $a$ as a `learning threshold' \cite{Tsodyks_1988},
\begin{equation}
J_{ij}=\frac{1}{C a}\sum_{\mu=1}^p\left(\xi_i^{\mu}-a\right)\left(\xi_j^{\mu}-a\right).\label{eq:oneshot}
\end{equation}
Note that this `classical' rule includes implausible positive contributions when both pre- and post-synaptic neurons are silent, and neglects a baseline value for synaptic weights, necessary to keep them positive excitatory weights. Both are simplifications convenient for the mathematical analysis, which have been discussed elsewhere (e.g., in \cite{Treves_1991}) and they will be assumed in the present model as well, though, as we will show, the first and more critical one will not be necessary once we introduce our modified rule. 

The above rule has been effectively used to store patterns drawn at random from the distribution with probability
\begin{equation}
P\left(\xi_i^{\mu}\right)=a\delta\left(\xi_i^{\mu}-1\right)+\left(1-a\right)\delta\left(\xi_i^{\mu}\right) \label{eq:trivial}
\end{equation}
independently for each unit $i$ and pattern $\mu$. In such conditions, the storage capacity of the network is $\alpha_c\propto a^{-1}$. This result assumes the limit of low sparseness, $a\ll 1$, which is the interesting case to model brain function, limit that we will also take in the rest of this paper.

Patterns that are correlated, unlike what is implied by the probability distribution in Eq.~\ref{eq:trivial}, cannot however be stored effectively in a network with weights given by Eq.~\ref{eq:oneshot}. For example, patterns intended to model correlated semantic memory representations have been considered for a long time `impossible to store' in an attractor network \cite{McRae_1997, McRae_1999, McRae_2006}.

\subsection{Network damage in the model}

Semantic impairments can result from damage of very diverse nature, like Herpes Encephalitis, brain abscess, anoxia, stroke, head injury and dementia of Alzheimer type, this last characterized by a progressive and widespread damage. How can we represent damage in our model network in a general way?

The model literature on attractor networks shows that the stability of memories depends on the parameter $\alpha=p/C$ as explained above, where $p$ can be considered in this case as fixed and equal to the number of concepts stored in the semantic memory of a patient. The sparseness $a$ also plays an important role, since the critical value of $\alpha$, or the storage capacity $\alpha_c$, varies inversely to $a$. In addition, we will show in this work that the distribution of popularity $a_i$ across neurons (the fraction of patterns in which each neuron $i$ is active) is a crucial determinant of the storage capacity when memories are correlated. However, it is interesting to notice that both in the modelling literature and in this paper, the total number of neurons in the network $N$ is not a determinant factor for the stability of memories, as long as it is large enough to apply statistics.

In our model, random damage to a memory network might affect only $C$ (if the damage is focalized on synapses) or $N$ and $C$ in the same proportion (if the damage is focalized on neurons), while the sparseness $a$ and the distribution of popularity (see below) should, to a first approximation, remain unchanged due to randomness. Since $N$ does not determine the stability of memories, here we simply model network damage as a decrease in the number of connections per neuron, $C$. Interestingly, forgetting in an intact network could be thought of as the modification of an increasing number of synaptic weights to values that are uncorrelated with the learned ones, and modeled in a similar way. The selective damage of an arbitrary group of synapses or neurons, instead, cannot be modelled simply as a decrease in $C$, and could lead to different and interesting results that are, however, outside the scope of this paper.

\section{Results}

\subsection{A rule for storing correlated distributions of patterns}

We consider a distribution of patterns in which Eq. \ref{eq:trivial} no longer applies, although, to simplify the analysis, we still assume patterns to have a fixed mean activity, as quantified by the sparseness $a$ (the more general case is treated in \cite{birmingham}, resulting in a more complicated analysis but no qualitative changes in the conclusions). We propose a learning rule similar to the one in Eq. \ref{eq:oneshot} with the variant that now learning thresholds are specific to each neuron,
\begin{equation}
J_{ij}=\frac{1}{C a}\sum_{\mu=1}^p\left(\xi_i^{\mu}-a_i^{post}\right)\left(\xi_j^{\mu}-a_j^{pre}\right). \label{eq:weightsII}
\end{equation}

Let us use a signal-to-noise analysis to identify appropriate values for such thresholds.
The field in Eq. \ref{eq:field} can be split into a signal and a noise part by assuming, without loss of generality, that pattern $1$ is being retrieved ($\sigma_j$ similar to $\xi_j^1$ for all $j$):
\begin{equation}
h_i=\frac{1}{C a}\left(\xi_i^{1}-a_i^{post}\right)\sum_{j=1}^N c_{ij}\left(\xi_j^{1}-a_j^{pre}\right)\sigma_j + \frac{1}{C a}\sum_{\mu=2}^p\left(\xi_i^{\mu}-a_i^{post}\right)\sum_{j=1}^N c_{ij}\left(\xi_j^{\mu}-a_j^{pre}\right)\sigma_j\label{eq:signaltonoise}
\end{equation}
where the first term in the RHS is the signal and the second term is the noise. As usual, the signal is a single macroscopic term that drives activity toward the desired attractor state, while a sum of many microscopic contributions comprises the noise. To analyze the latter we assume that $\xi_i^{\mu}$ and $\xi_j^{\mu}$ are statistically independent variables, as long as $i\neq j$ (whereas we \textit{do not} require $\xi_i^{\mu}$ and $\xi_i^{\nu}$ to be independent; on the contrary, the aim is to handle their correlation). If this condition of independence among units, which is central to our analysis, is fulfilled, the noise term can be viewed, to a first approximation, as generated by a gaussian distribution with mean
\begin{equation}
\ll noise \gg =\frac{p-1}{C a}\sum_{j=1}^N c_{ij}\sigma_j \left(\ll \xi_i^{\mu}\gg_{\mu} -a_i^{post}\right)\left(\ll \xi_j^{\mu}\gg_{\mu} -a_j^{pre}\right).\label{eq:mean}
\end{equation}
If this mean is different from zero, the noise scales up with $p$, which is the first cause of the performance collapse mentioned above (the optimal one-shot learning rule for uncorrelated patterns has $a_k^{post}=a_k^{pre}=a$ for all $k$, which results in general in a mean noise different from $0$). For $\ll noise \gg $ in Eq. \ref{eq:mean} to vanish, at least to leading order in $p$, we must choose either $a_i^{post} = \ll \xi_i^{\mu}\gg_{\mu}$ or $a_j^{pre} = \ll \xi_j^{\mu}\gg_{\mu}$. We choose the latter

\begin{equation}
a_i^{pre}= a_i\equiv \frac{1}{p}\sum_{\mu=1}^p \xi_i^{\mu}\label{eq:popularity}
\end{equation}
where we have introduced $0\leq a_i\leq 1$, the $popularity$ of neuron $i$, that measures how shared is the activity of this neuron among the patterns in memory. Once this particular choice has been made, one sees from Eq. \ref{eq:weightsII}  that the contribution of $a_i^{post}$ to the field $h_i$ vanishes, and its exact value is irrelevant. We then choose $a_i^{post} = 0$ for all $i$.

The next step is to analyze how the variance of the noise distribution scales up with $p$ and $C$. We have
\begin{equation}
\ll (noise-\ll noise\gg )^2\gg  = \frac{1}{C^2 a^2}\sum_{\mu,\nu=2}^p\xi_i^{\mu}\xi_i^{\nu}\sum_{j,k=1}^N c_{ij}c_{ik}\sigma_j \sigma_k\left(\xi_j^{\mu} -a_j\right)\left(\xi_k^{\nu} -a_k\right)\nonumber
\end{equation}
which can be divided into four contributions that scale differently with $p$ and $C$, depending on whether or not $j$ and $k$ on one side and $\mu$ and $\nu$ on the other are equal:
\begin{eqnarray}
\ll (noise-\ll noise\gg )^2\gg = & &\frac{1}{C^2 a^2}\sum_{\mu=2}^p\xi_i^{\mu}\sum_{j=1}^N c_{ij}\sigma_j^2\left(\xi_j^{\mu} -a_j\right)^2+  \nonumber\\
& + &\frac{1}{C^2 a^2}\sum_{\mu\neq\nu=2}^p\xi_i^{\mu}\xi_i^{\nu}\sum_{j=1}^N c_{ij}\sigma_j^2\left(\xi_j^{\mu} -a_j\right)\left(\xi_j^{\nu} -a_j\right)+\nonumber\\
& + &\frac{1}{C^2 a^2}\sum_{\mu=2}^p\xi_i^{\mu}\sum_{j\neq k=1}^N c_{ij}c_{ik}\sigma_j \sigma_k\left(\xi_j^{\mu} -a_j\right)\left(\xi_k^{\mu} -a_k\right)+\nonumber\\
& + &\frac{1}{C^2 a^2}\sum_{\mu\neq\nu=2}^p\xi_i^{\mu}\xi_i^{\nu}\sum_{j\neq k=1}^N c_{ij}c_{ik}\sigma_j \sigma_k\left(\xi_j^{\mu} -a_j\right)\left(\xi_k^{\nu} -a_k\right).
\label{eq:variance4}
\end{eqnarray}

The first term in the RHS scales like $(p-1)/C\simeq\alpha$, the second one like $(p-1)(p-2)/C$, the third one like $(p-1)$ and the fourth like $(p-1)(p-2)$. Remembering, however, our definition of popularity in Eq. \ref{eq:popularity}, and the statistical independence between neurons, one can see that the leading contributions to the second to fourth term vanish. The remaining dependency of the variance on $\alpha$ is similar to the one found in classical models of autoassociative memory with independent or randomly correlated patterns, indicating that the new rule
\begin{equation}
J_{ij}=\frac{1}{C a}\sum_{\mu=1}^p \xi_i^{\mu}\left(\xi_j^{\mu}-a_j\right) \label{eq:weights}
\end{equation}
is a generalization of the Hopfield model appropriate to the storage of correlated patterns.

\begin{figure}[ht]
\centerline{\hbox{\epsfig{figure=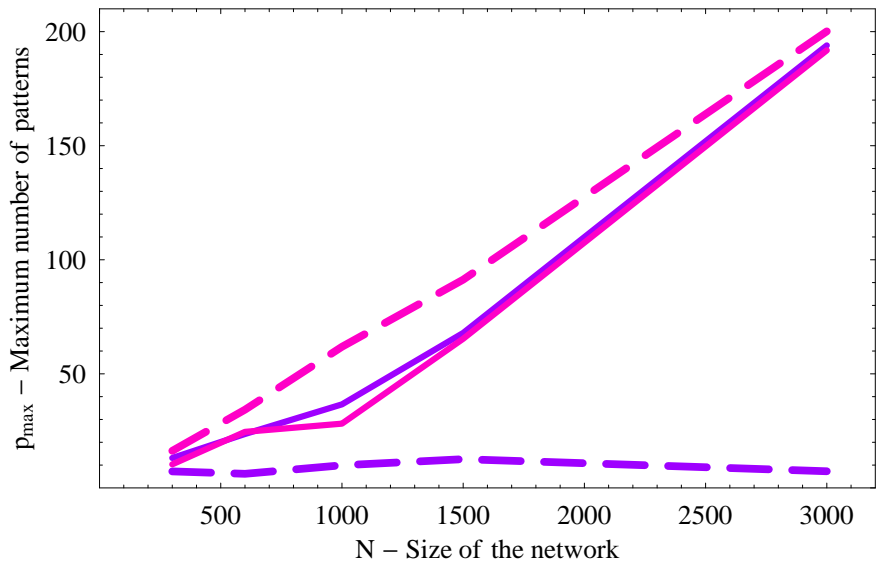,width=12cm,angle=0}}}
\caption{The critical value $p_{max}$ measured as the value of $p$ at which $70\%$ of the patterns are retrieved succesfully. We show $p_{max}$ as a function of $N$ using the proportion $C=0.17 N$ for the four combinations of two learning rules and two types of dataset. Violet: one shot `standard' learning rule of Eq. \ref{eq:weightsII}. Pink: modified rule of Eq. \ref{eq:weights}. Solid: trivial distribution of randomly correlated patterns obtained from Eq. \ref{eq:trivial}. Dashed: non-trivially correlated patterns obtained using a hierarchical algorithm. In three cases the scaling of $p_{max}$ with $C$ is linear, as in the classical result. Only in the case of one-shot learning of correlated patterns there is a storage collapse.}
\label{fig:scaling}
\end{figure}

Figure \ref{fig:scaling} shows simulations of networks of different size and connectivity, employing 
either the classical or our modified learning rule, to store either uncorrelated or correlated memories, as described in \textit{Methods}. The hierarchical algorithm described in \cite{NACO} allows us to construct datasets of different $p$ and $N$ values with approximately the same correlation statistics. The four curves result from the combination of the two different learning rules, the standard rule in Eq. \ref{eq:oneshot} and the one in Eq. \ref{eq:weights}, with two types of pattern distribution, correlated or not. With the standard, one-shot learning rule, the number of uncorrelated patterns constructed using Eq. \ref{eq:trivial} that can be stored and correctly retrieved, $p_{max}$, grows linearly with 
the connectivity $C$. With non-trivial correlations among patterns, however, the storage capacity collapses: rather than scaling linearly with $C$, $p_{max}$ even decreases toward $0$ for very high values of $C$. This catastrophe is reversed when the popularity $a_i$ replaces the sparseness $a$ as a learning threshold, bringing $p_{max}$ back to its usual linear dependence on $C$. The linear dependence of course holds also when the more advanced rule is applied to the original dataset of uncorrelated (i.e., randomly correlated) patterns. Finally, it is important to note that the success in retrieving patterns stored with the rule of Eq. \ref{eq:weights} does not depend on the algorithm that we used to construct the patterns, but rather shows the generality of the rule, as we do not include in it information about how patterns are constructed. We have tested the modified network with other sets of patterns (such as the random patterns in the same Figure or those described in \textit{Methods}: patterns resulting from setting arbitrary popularity distributions across neurons as shown in Figure \ref{fig:pfinal3} or patterns taken from the semantic feature norms of McRae and colleagues \cite{birmingham, Mcrae_2005}) always reaching levels of retrieval that are consistent with the predictions of the theory.

Having defined the optimal model for the storage of correlated memories, we analyze in the following sections the storage properties and its consequences through mean field equations. We note that the average of the popularity across neurons is $\sum_{j=0}^Na_j/N=a\ll 1$. In the interesting limit we will consider the popularity $a_i$ generally near $0$, and only exceptionally close to $1$. 

\subsection{Retrieval with no interference: $\alpha\simeq 0$}

If a pattern is being retrieved in a network with very low memory load ($\alpha \simeq 0$), the interference due to the storage of other patterns is negligible. The field in Eq. \ref{eq:field} is driven by a single term corresponding to the contribution of the pattern that is being tested for retrieval (which we call pattern $1$), or, in other words, the signal term,
\begin{equation}
h_i\simeq \xi_i^{1}\left[\frac{1}{C a}\sum_{j=1}^N c_{ij}\left(\xi_j^{1}-a_j\right)\sigma_j\right].\label{eq:field2}
\end{equation}
This can be re-expressed by defining the variables
\begin{equation}
m_i^{\mu}\equiv \frac{1}{C a}\sum_{j=1}^N c_{ij}\left(\xi_j^{\mu}-a_j\right)\sigma_j
\end{equation}
and by noticing that, since $N$ and $C$ are large (in the thermodynamic limit both tend to infinity) and $c_{ij}$ is a random connectivity matrix,
\begin{equation}
m_i^1\rightarrow m\equiv \frac{1}{Na}\sum_{j=1}^N\left(\xi_j^1-a_j\right)\sigma_j\label{eq:m},
\end{equation}
that is, the average of $(\xi_j^1-a_j)\sigma_j$ across neurons. The variable $m$ always refers to the pattern that is being tested for retrieval, and it measures its overlap with the state of the network. 

Inserting Eq. \ref{eq:m} into Eq. \ref{eq:field2} we obtain
\begin{equation}
h_i\simeq \xi_i^{1}m.
\end{equation}
This expression can be inserted into Eq. \ref{eq:activation} to obtain the updated value of $\sigma_j$ for all neurons $j=1\dots N$. If the state of the network is stable, $\sigma_j$ does not change with updating, so it can be reinserted into Eq. \ref{eq:m}, yielding a single equation that describes the stable attractor states of the system
\begin{equation}
m=\frac{1}{Na}\sum_{j=1}^N\left(\xi_j^1-a_j\right)\left[1+\exp \beta \left(U-\xi_j^1 m \right)\right]^{-1}\label{eq:m2}.
\end{equation}

Splitting the sum into the $aN$ terms in which $\xi_j^1=1$ and the $(1-a)N$ terms in which $\xi_j^1=0$, we can rewrite it as
\begin{equation}
m=\left(1-a^1\right)\{[1+\exp \beta \left(U-m \right)]^{-1}-[1+\exp \beta U]^{-1}\}\label{eq:alfa0}
\end{equation}
where the new parameter $0\leq a^{\mu}\leq 1$ can be thought of either as the average popularity of the neurons active in pattern $\mu$ or as the average overlap between pattern $\mu$ and the other patterns:
\begin{equation}
a^{\mu}\equiv \frac{1}{Na}\sum_{j=1}^N \xi_j^{\mu} a_j=\frac{1}{p}\sum_{\nu=1}^p\left[\frac{1}{Na}\sum_{j=1}^N \xi_j^{\mu} \xi_j^{\nu}\right].\label{eq:kappa}
\end{equation}
Note that for the interesting limit of very sparse activity, in most cases $a^{\mu}\ll 1$. From the definition of $m$ in Eq. \ref{eq:m} it can be noted that $m=1-a^1\simeq 1$ for perfect retrieval (i.e., $\{\sigma_j\}\equiv \{\xi_j^1\}$) and $m=a-a^{\sigma}\simeq 0$ if the activity $\sigma$ of the network has sparseness $a$ but is unrelated to $\xi^1$, i.e., retrieval fails.

\begin{figure}[ht]
\centerline{\hbox{\epsfig{figure=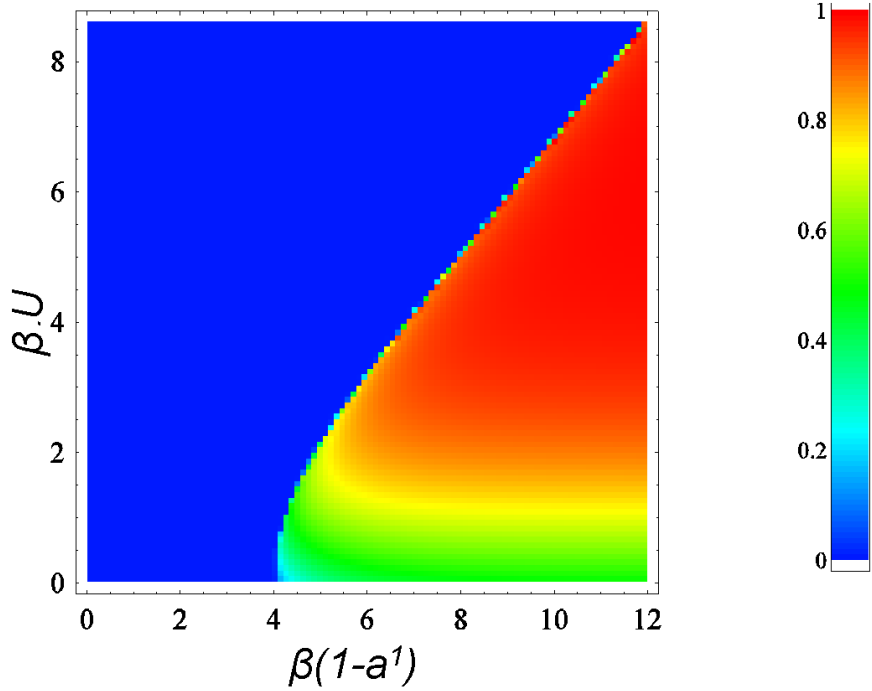,width=12cm,angle=0}}}
\caption{Numerical solutions of Eq. \ref{eq:alfa0} varying the two relevant parameters: $\beta (1-a^1)$ on the $x$ axis and $\beta U$ on the $y$ axis. A first order phase transition is observed in the value of $m$ that solves Eq. \ref{eq:alfa0}. In the limit $\beta \rightarrow \infty $ the transition occurs along the identity line $1-a^1=U$.}
\label{fig:phase}
\end{figure}

Eq. \ref{eq:alfa0} always admits the solution $m=0$, and it may have another stable solution depending on two combinations of parameters: $\beta U$ and $\beta (1-a^1)$. Whenever this non-zero solution exists, retrieval is possible. In Figure \ref{fig:phase} we show, as a function of the two parameters, the highest value of $m$ that solves Eq. \ref{eq:alfa0}. A first order phase transition is observed: given a fixed value of $\beta U$ there is a critical value of $\beta (1-a^1)$ below which the only solution to Eq. \ref{eq:alfa0} is $m=0$, i.e., no retrieval. In the `zero-temperature' ($\beta\to \infty$) limit, the condition for the existence of a non-zero solution in Eq. \ref{eq:alfa0} reduces to $m=(1-a^1)\geq U$, showing that at the critical point $a^1_c=1-U$. Clearly, the choice $U=0$ would permit the retrieval of patterns with arbitrary values of $a^1$ (which is, by definition, not larger than $1$), but as shown in \cite{Buhmann_1989, Tsodyks_1988} and in the following sections, a threshold value of order $1$ is necessary to obtain an extensive storage capacity, close to optimal, when interference due to the storage of other patterns is not negligible. 

An intuitive explanation of Figure \ref{fig:phase} would be the following. The learning rule in Eq. \ref{eq:weights} implies that the network is less \textit{confident} of any neuron $j$ with high popularity, since its positive contributions to outgoing weights are proportional to $1-a_j$. This implies that the more popular is, on average, the ensemble of neurons underlying a given memory (as expressed by its $a^1$ value), the less able it is to sustain, through neural activity, the corresponding attractor state. When the average activating signal is smaller than the threshold $U$, retrieval is no longer possible.

\subsection{Retrieval with interference: diluted networks}

To treat the case of extensive storage, $p$ scaling up with $C$, we consider the so called {\em highly diluted}
approximation, which is valid when either $C\ll N$ (`diluted', i.e. sparse connectivity proper, \cite{Gardner_1987}) or $a\ll 1$ 
(very sparse activity, \cite{Treves_1991}). There are two independent motivations to study such a limit: on one side it approximates real cortical networks, with their sparse connectivity and sparse firing, on the other, calculations are much simpler than for fully connected networks, enabling deeper analysis and wider generalization. In addition, one obtains in this limit differential equations for the dynamical evolution of all relevant variables, valid also outside of equilibrium \cite{Gardner_1987}. Such an approach is outside the scope of this paper, and it is left for future studies. It is worth mentioning that some experimental work on semantic memory \cite{Sartori_2004, Sartori_2005} is based on a dynamical view of the networks involved in semantic processing, as it focuses on the type of input cues that can lead to successful retrieval. 

The highly diluted approximation takes into account in the field $h_i$ a signal term and a gaussian noise, while neglecting the effect of a second source of noise due to the propagation of neural activity around closed loops of synaptic connections. These effects scale in general like $\alpha a C/N$ \cite{yasser_2004, birmingham}, and are therefore negligible as $C/N\to 0$, $a\ll 1$ or, as in the previous section, $\alpha \simeq 0$. 

In Eq. \ref{eq:variance4} we had already obtained an expression of the variance of the noise part of the field $h_i$ when considering it to be purely gaussian. After computing the average over $\mu$ in the surviving first term, we obtain
\begin{equation}
\ll (noise-\ll noise\gg )^2\gg=\alpha\ a_i\left[\frac{1}{C a^2}\sum_{j=1}^N c_{ij}a_j\left(1-a_j\right)\sigma_j^2\right].
\end{equation}
The expression between square brackets depends on $i$ only through the connectivity matrix $c_{ij}$. As in Eq. \ref{eq:m}, we can take advantage of the fact that $c_{ij}$ is random and $C$ large, and replace the sum with an average over all neurons. We can conclude that $\ll (noise-\ll noise\gg )^2\gg=\alpha a_i q$, where we define
\begin{equation}
q\equiv \frac{1}{N a^2}\sum_{j=1}^N a_j\left(1-a_j\right)\sigma_j^2.\label{eq:q}
\end{equation}

The local field then becomes
\begin{equation}
h_i=\xi_i^1 m+\sqrt{\alpha a_i q} z_i
\end{equation}
where $z_i$ may be assumed to be drawn from a normal distribution with mean $0$ and variance $1$, statistically independent with all other variables \footnote{In the simplest signal-to-noise approach \cite{potts-2005} two `worst-case' conditions must be met in order to have stable attractors: $h_i=m -\sqrt{variance}>U$ for values of $i$ in which $\xi_i^1=1$ and $h_i=\sqrt{variance}<U$ for $\xi_i^1=0$. This shows that the optimal value of $U$ is $m/2\simeq (1-a^{\mu})/2$, which depends on global rather than local information. Interesting corrections in which the optimal value of $U$ depends on $a_i$ and is thus different for each neuron might come out of considering the non-diluted case, including an additional term in the local field $h_i$ as mentioned above.}. To describe attractors of the system, as previously, we insert the field into Eq. \ref{eq:activation} to obtain the stable value of $\sigma_j$, which can be re-inserted into the definition of $m$ in Eq. \ref{eq:m},
\begin{equation}
m=\frac{1}{Na}\sum_{j=1}^N\left(\xi_j^1-a_j\right) \left[1+\exp \beta \left(U-\xi_j^1 m -\sqrt{\alpha a_j q} z_j\right)\right]^{-1}.\label{eq:step1}
\end{equation}
Making use of the independence of $z_j$ with respect to $a_j$ and $\xi_j^1$, we can  take its average. The highly diluted version of Eq. \ref{eq:m2} is then
\begin{equation}
m=\frac{1}{Na}\sum_{j=1}^N\left(\xi_j^1-a_j\right) \int_{-\infty }^{\infty}Dz \left[1+\exp \beta \left(U-\xi_j^1 m -\sqrt{\alpha a_j q} z\right)\right]^{-1}
\end{equation}
where the gaussian differential is
\begin{equation}
Dz\equiv dz \frac{1}{\sqrt{2\pi}}\exp\left(-\frac{z^2}{2}\right)\label{eq:gauss}
\end{equation}
expressing the distribution of $z_j$.

In the following, for simplicity, we will take the limit of zero temperature, $\beta \rightarrow \infty $. The equation for $m$ becomes
\begin{equation}
m=\frac{1}{Na}\sum_{j=1}^N\left(\xi_j^1-a_j\right) \phi\left(\frac{\xi_j^1 m-U}{\sqrt{\alpha a_j q}}\right) \label{eq:mdil}
\end{equation}
where
\begin{equation}
\phi \left(y\right)=\frac{1}{2}\left(1+{\rm erf} \left(\frac{y}{\sqrt{2}}\right)\right)
\end{equation}
is a sigmoidal function increasing monotonically from $0$ to $1$, with $\phi(0)=1/2$. Since in Eq. \ref{eq:mdil} the terms are not linear in $a_j$, it is not straightforward to obtain the new version of Eq. \ref{eq:alfa0}. To do so we must first introduce the distribution of popularity across neurons, given by the probability
\begin{equation}
F\left(x\right)\equiv P\left(a_j=x\right),
\end{equation}
and the distribution of popularity across neurons that are active in the pattern we are testing for retrieval,
\begin{equation}
f\left(x\right)\equiv P\left(a_j=x|\xi_j^{1}=1\right).
\end{equation}
The purpose of introducing these distributions is to convert a discrete set of popularities $\{a_j\}$ into a continuous distribution, where the popularity is represented by the variable $x$. Since $N$ is large, we can transform the sum in Eq. \ref{eq:mdil} into an integral over these distributions. As a result we obtain the equation
\begin{eqnarray}
m&=&\int_0^1 dx f(x)\left\{(1-x)\phi \left(\frac{m-U}{\sqrt{\alpha x q}}\right)+x \phi \left(\frac{-U}{\sqrt{\alpha x q}} \right)\right\}-\nonumber\\
& &-\frac{1}{a}\int_0^1 dx F(x)x\phi \left(\frac{-U}{\sqrt{\alpha x q}}\right)\label{eq:mfinal},
\end{eqnarray}
which extends Eq. \ref{eq:alfa0} to the case of non negligible interference.

Since this equation depends not only on $m$ but also on $q$, we need a second equation to close the system and univocally describe the stable states of the network. From the definition of $q$ in Eq. \ref{eq:q} we can repeat the steps \ref{eq:step1} to \ref{eq:mdil} and obtain, for stable states and in the limit of zero temperature,
\begin{equation}
q= \frac{1}{N a^2}\sum_{j=1}^N a_j\left(1-a_j\right)\left[\phi\left(\frac{\xi_j^1 m-U}{\sqrt{\alpha a_j q}}\right)\right]^2.\label{eq:qmedio}
\end{equation}
Introducing again the distributions of popularity -- steps \ref{eq:mdil} to \ref{eq:mfinal} -- we can simplify this expression into
\begin{eqnarray}
q&=&\frac{1}{a}\int_0^1 dx f(x) x(1-x)\left\{\phi \left(\frac{m-U}{\sqrt{\alpha x q}}\right)- \phi \left(\frac{-U}{\sqrt{\alpha x q}} \right)\right\}+\nonumber\\
& &+\frac{1}{a^2}\int_0^1 dx F(x)x(1-x)\phi \left(\frac{-U}{\sqrt{\alpha x q}}\right).\label{eq:qfinal}
\end{eqnarray}

Eqs. \ref{eq:mfinal} and \ref{eq:qfinal} describe the stable states of the network in this `diluted' approximation. As in the noiseless case, a phase transition separates regions of parameter space where a solution with $m\sim 1-a^1$ exists from regions where the only solution is $m=q=0$. The latter can now be reached by increasing $\alpha= p/C$, i.e. the memory load. In other words, the phase transition to no retrieval determines the storage capacity of the system. If $f(x)=F(x)=\delta(x-a)$, which is the case for uncorrelated patterns, the classical equations for highly diluted binary networks \cite{Buhmann_1989, Tsodyks_1988} are re-obtained, and the critical value of the memory load scales like
\begin{equation}
\alpha_c\propto \frac{1}{a \ln(1/a)}\label{eq:tsodyks}
\end{equation}
for the relevant sparse limit $a\ll 1$.

How does this classical result generalize to the case of correlated representations?

\subsection{The storage capacity}

Already at first glance, the system of Eqs. \ref{eq:mfinal} and \ref{eq:qfinal}, which determine the storage capacity of a network with correlated patterns, reveals a new property of associative memories. In both equations, the second term in the RHS depends on $F(x)$ and is thus common to the retrieval of any pattern. However, the RHS of both equations depends also on $f(x)$, the distribution of popularity among neurons active in the pattern that is being retrieved. In the general case, this distribution is different for every pattern, so that \textit{the stability properties of the associated attractors will differ from pattern to pattern}. 

To understand this idea it is convenient to think about the storage capacity as $p/C_{min}$ (the minimum connectivity necessary to sustain retrieval) rather than as $p_{max}/C$ (the maximum number of patterns that can be stored). In this view, each of $p$ memory states stored in a network has an associated value of $C_{min}$ that depends on its own statistical properties and on the statistical properties of the whole dataset. Any particular pattern can be retrieved only if the actual connectivity level $C$ is higher than the value of $C_{min}$ associated to it. 

This view is of particular interest to analyze category specific deficits in semantic memory. We can think of $p$ as being relatively fixed, corresponding, in the model, roughly to all the concepts acquired by a healthy subject during an entire life. A mild and non-selective damage of the network might decrease the parameter $C$, which would selectively affect the memories with a high value of $C_{min}$, while sparing the others.

\subsubsection{An entropy characterization of the noise}
To analyze Eqs. \ref{eq:mfinal} and \ref{eq:qfinal} we first consider that $\alpha$ and $U$ are small enough to ensure that the retrieval is possible and that $\phi \left(\frac{m-U}{\sqrt{\alpha x q}}\right)\sim 1$ and $\phi \left(\frac{-U}{\sqrt{\alpha x q}}\right)\sim 0$. Following this, any pattern that we choose to test for retrieval has $m\simeq 1-a^1$, as we had found for $\alpha \simeq 0$ and a value of the noise variable $q$ that is proportional to the average of $a_j(1-a_j)$ over the neurons that are active in the pattern (as can be seen from Eqs. \ref{eq:qmedio} or \ref{eq:qfinal}), or in other words,
\begin{equation}
S_f\equiv \int_0^1 x(1-x)f(x).\label{eq:entropy}
\end{equation}
Similarly to Shannon's entropy, $S_f$, and in consequence the noise variable $q$, approaches $0$ if neurons in the distribution are all either very popular or unpopular in their firing, while it is maximum ($S_f=1/4$) when $f(x)=\delta(x-1/2)$, i.e. all neurons have popularity $a_i=1/2$ \footnote{Technically, this function applied to a single unit is Tsallis' entropy with parameter $q=2$. Note, however, that Tsallis' entropy is not additive for independent events, while our $S_f$ is clearly a normalized extensive quantity.}. Thus, a pattern will be better retrieved if a) it includes as unpopular neurons as possible (as shown previously, to ensure $m=1-a^1>U$) and b) its neurons have a low `entropy' value $S_f$, in order to minimize the noise $q\simeq S_f/a$.

An intuitive explanation of this comes from the analysis of the influence of neuron $j$ as noise in the field $h_i$, proportional to $\sum_{\mu\neq1} \xi_i^{\mu}(\xi_j^{\mu}-a_j)$ as shown in Eq. \ref{eq:signaltonoise}. If the popularity of neuron $j$ is very low, terms of this noise where $\xi_j^{\mu}=1$ are large contributions (proportional to $1-a_j$), but very infrequent, while terms in which $\xi_j^{\mu}=0$ are very frequent but only proportional to $a_j\ll 1$. The exact opposite pattern emerges if neuron $j$ is very popular. As a result of this, in both cases the noise is very low. In the extreme of $a_j=0$ or $a_j=1$ the noise is exactly zero, since contributions of order $1$ occur with probability $0$ and inversely. In such a case the dynamics of the network is guided purely by the signal terms, that take $h_i$ toward the correct value for retrieval. The case in which the noise is maximal is when the probability of neuron $j$ to be active is $a_j=1/2$ and each term of the contribution of neuron $j$ to the noise in the field $h_i$ is proportional to $1-a_j=1/2$ or $a_j=1/2$. Finally, since the noise is also proportional to $\sigma_j$ and pattern $1$ is being retrieved, this effect is important only for the neurons $j$ that are active in this pattern, explaining fully Eq. \ref{eq:entropy}.

\subsubsection{The storage capacity is inverse to $S_f$}
As $\alpha$ increases, the assumption $\phi \left[(m-U)/\sqrt{\alpha x q}\right]\sim 1$ becomes eventually incorrect and for some critical value $\alpha_c$ a retrieval solution with $m\sim 1-a^1$ no longer exists. A generally fair approximation when studying storage capacity is to assume that $\alpha_c$ scales inversely to the factor that accompanies $\alpha$ in the argument of $\phi$, which in this case is $xq$. However, since $x$ is a variable that spans the whole range from $0$ to $1$, the approximation is not useful in itself. In more general terms, $\alpha_c$ should scale inversely to $x_fq$, with $0<x_f<1$ some intermediate value with a strong dependence on $f(x)$. In this section we consider the case in which the variance of $F(x)$ is small enough to allow the approximation of $x$ by its average $a$ in the argument of $\phi$, while in \textit{Methods} we analyze some more general examples. 

Our first order approximation, assuming $\alpha$ inverse to $aq$ and $q\simeq S_f/a$, leads to

\begin{equation}
\alpha_c\propto \frac{1}{S_f}.\label{eq:alpha}
\end{equation}
In line with what we had explained intuitively, the storage capacity, or $C_{min}/p$, is inverse to the entropy $S_f$ of the pattern. In the classical case of randomly correlated patterns $S_f=a(1-a)\sim a$ (again, assuming cortical activity to be sparse, the interesting approximation is always $a\ll 1$), which leads to the Tsodyks and Feigel'man result in Eq. \ref{eq:tsodyks}, without the logarithmic correction. 

This correction appears only when $\phi \left(-U/\sqrt{\alpha a q}\right)$ starts to be significantly different from $0$. The largest contribution is the one given by the second term in the RHS of Eq. \ref{eq:qfinal}, since it is not negligible when $\phi \left(-U/\sqrt{\alpha a q}\right)$ is of order $a$ (considering $a\ll 1$), while the other neglected terms are only relevant when $\phi \left(-U/\sqrt{\alpha a q}\right)$ is of order $1$. Again, we use the approximation of low variance, so the term we are interested in becomes
\begin{equation}
\mathcal{T}_2=\frac{1}{a^2}\phi \left(\frac{-U}{\sqrt{\alpha a q}}\right)\int_0^1 dx F(x)x(1-x)\equiv \frac{1}{a^2}\phi \left(\frac{-U}{\sqrt{\alpha a q}}\right)S_F,
\end{equation}
where, similarly to $S_f$, we define $S_F$ as the entropy of the distribution $F(x)$. This term is near $0$ for very small values of $\alpha$, where $q$ is dominated by the first term of Eq. \ref{eq:qfinal}, which can still be considered as $S_f/a$, and it becomes significant only when both terms are of comparable magnitude. If this happens at values of $\alpha$ that are smaller than the one indicated by Eq. \ref{eq:alpha}, the correction introduced by this term is relevant. To estimate this correction we impose the first and second terms of Eq. \ref{eq:qfinal} to be about equal ($\mathcal{T}_2\simeq S_f/a$) and consider $a\ll 1$, which leads to

\begin{equation}
\phi \left(-\frac{U}{\sqrt{\alpha_c S_f}} \right)\simeq \frac{a S_f}{S_F}.\label{eq:phi}
\end{equation}

Inverting the function $\phi$ we obtain $\alpha_c$ as

\begin{equation}
\alpha_c\simeq\frac{1}{2S_f}\left[ \frac{U}{{\rm erf}^{-1} \left(1-\frac{2aS_f}{S_F} \right)} \right]^2.
\end{equation}
The inverse error function can be approximated as
\begin{equation}
{\rm erf}^{-1}(1-y)\sim \sqrt{\ln\left( \sqrt{\frac{2}{\pi}}\frac{1}{y}\right)}
\end{equation}
for small values of $y$. Since $F(x)$ has low variance, $S_f, S_F \sim  a \ll 1$ and $a S_f/S_F$ can be taken to be a small quantity. We then approximate
\begin{equation}
\alpha_c\simeq\frac{1}{2S_f}\left[ \frac{U^2}{\ln\left( \frac{S_F}{\sqrt{2\pi}aS_f}\right)} \right]\propto  \frac{1}{S_f \ln\left( \frac{S_F}{aS_f}\right)}.\label{eq:info}
\end{equation}
If this scaling of $\alpha_c$ is lower than indicated by Eq. \ref{eq:alpha} (or, in other words, if $\ln( S_F/(aS_f))>1$) this correction is relevant. Finally, in the case of trivial correlations $f(x)=F(x)=\delta(x-a)$ and consequently $S_f=S_F\simeq a$. The full classical result of Eq. \ref{eq:tsodyks} is then reproduced by Eq. \ref{eq:info}, indicating that the latter is a generalization of the former. 
\comment{\begin{equation}
S_g\equiv \int_0^1 x(1-x)g(x)\label{eq:i}
\end{equation}
where $g(x)$ is any distribution of popularity values (in particular, either $f(x)$ or $F(x)$, resulting in $S_f$ and $S_F$). Similarly to Shannon's entropy, $S_g$ approaches $0$ if neurons in the distribution are all either very popular or unpopular in their firing, while it is maximum ($S_g=1/4$) when $g(x)=\delta(x-1/2)$, i.e. all neurons have popularity $a_i=1/2$. }

\begin{figure}[ht]
\centerline{\hbox{\epsfig{figure=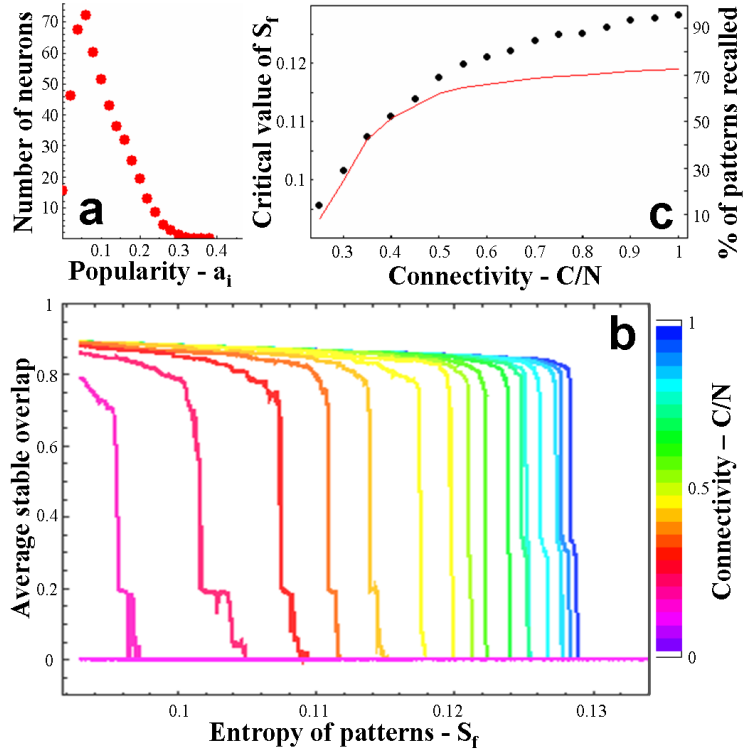,width=10cm,angle=0}}}
\caption{Simulations of the storage capacity of a network storing patterns with an arbitrary correlation distribution F(x). The parameters are $N=500$, $p=50$, $a=0.1$, $U=0.35$ and variable $C$. For all values of $C$ each pattern is tested $10$ times for stability, with different connectivity matrices $c_{ij}$. \textbf{a} Popularity distribution across the whole network, $F(x)$. Note that neurons with $a_i=0$ do not really participate in network dynamics, making the effective values of $C$ and $N$ slightly lower. \textbf{b} Stable value of $m$ for each pattern vs. its $S_f$ value. The data has been smoothed by taking the median over a moving window. From blue toward violet: connectivity $C/N$ starting with $1$ and decreasing in steps of $0.05$. For each color, the graph shows that some patterns are retrieved while others are not, corresponding to low and high values of $S_f$. The critical value of $S_f$ at which the transition occurs moves to the left as the connectivity is reduced, which, as explained in the Introduction, is the strongest effect of random network damage. \textbf{c} Storage capacity computed from the step-like transitions in \textbf{b}. Black dots, left axis: critical value of $S_f$ vs. connectivity, showing the maximum retrievable $S_f$ supported by the $C$ connections of the network. Red line, right axis: percent of patterns with a value of $S_f$ lower than the critical one.}
\label{fig:pfinal3}
\end{figure}

In \textit{Methods} we find expressions similar to \ref{eq:info} for wider distributions of $F(x)$. As we show, the slower the decay of the tail of a smooth distribution $F(x)$ with increasing $x$, the poorer is performance in terms of storage capacity. If the decay of $F(x)$ is exponential or faster, the $1/S_f$ scaling of Eq. \ref{eq:info} holds with at most a larger logarithmic correction. If the decay is a power-law, instead, the scaling is much poorer: $\alpha_c\propto a/S_f$, with, as usual, $a\ll 1$. 

\begin{figure}[ht]
\centerline{\hbox{\epsfig{figure=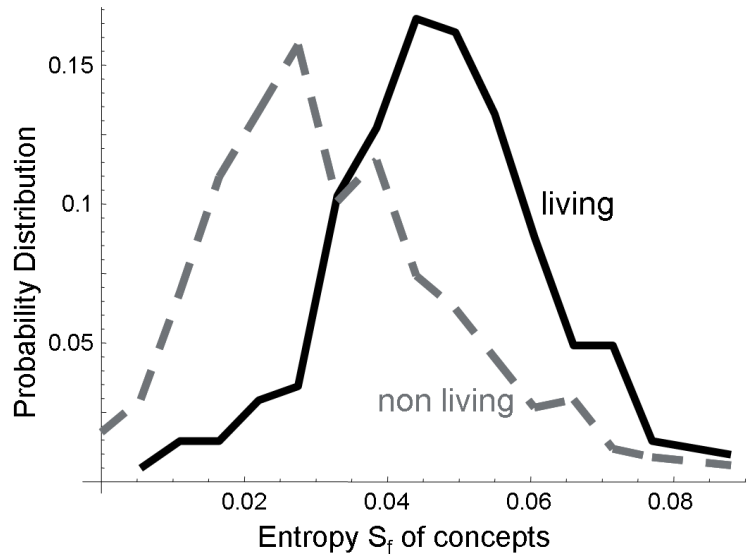,width=12cm,angle=0}}}
\caption{Distribution of $S_f$ in concepts belonging to the `living' and the `non living' categories obtained from the feature norms of McRae and colleagues \cite{McRae_norms}. Living things have a distribution centered at higher values of $S_f$, which in terms of our analysis means that they are more informative but also more susceptible to damage, as observed in patient studies.}
\label{fig:mcrae}
\end{figure}

\subsubsection{Informative memories are less robust}

In Figure \ref{fig:pfinal3} we show results of simulations using a distribution of correlated patterns (see details in \textit{Methods}), focusing on how the successful retrieval of a pattern depends on its $S_f$ value, and how a decrease in $C$ results in the selective lost of memories. This illustrates how the effective memory load of a network depends not only on the number of patterns that are being stored but also on how \textit{informative} they are. An autoassociative memory could store virtually infinite patterns, for example, if they were constructed in such a way that all of the neurons contributed vanishing entropy, and hence were minimally informative: this would be the case if some neurons were active in nearly every pattern, while others in none, keeping the mean activity fixed to a value $a$. This result is in agreement with the notion that any associative memory network is ultimately constrained in the amount of information each of its synapses may store \cite{Gardner_1988}.

The other interesting aspect of Eqs. \ref{eq:mfinal} and \ref{eq:qfinal} is that memory patterns are rather independent from one another in their retrievability. In the process of lowering $C$ (which is, as explained in \textit{Introduction}, the strongest effect of random network damage in our model) any pattern with a low value of $S_f$ would be retrieved even when most of the other patterns have become irretrievable. Generally speaking, \textit{informative memories are lost, while non-informative ones are kept}.

This model thus offers a quantitative explanation of category specific effects, along principles similar to those suggested, in a non mechanistic way, by several previous studies  \cite{Tyler_2000, Sartori_2004, McRae_1997}. In our network, the classical dichotomy would be verified if the semantic representations of $living$ things had on average higher values of $S_f$ than those of $non living$ things, a plausible assumption that can be assessed using evidence in the relevant literature. As an example, we analyze the feature norms of McRae and colleagues, experimentally obtained representations of $541$ concepts in terms of $2526$ features \cite{McRae_norms} (see \textit{Methods}). In Figure \ref{fig:mcrae} we show that the distributions of $S_f$ in the two categories overlap, but they are centered around different values of $S_f$, with living things on average more informative, hence more vulnerable to damage -- a trend that is consistent with our analysis \footnote{One could feel tempted to store the patterns obtained from these norms in a network in order to simulate damage in a more direct way. Some new technical problems arise, however, since the sparseness $a$ is not constant across patterns. In addition, the performance of the network is very poor due to the fact that the popularity distribution of the norms $F(x)$ has a power-law decay. This poor performance does not contradict the theory developed here, but rather validates it, as elaborated in \cite{birmingham}.}.

\section{Discussion}
Several experimental studies investigating semantic memory from the perspective of feature representation suggest that the representation of concepts in the human brain present non-trivial correlations  \cite{Vigliocco_2002, Garrard_2001}, presumably reflecting to some extent non-trivial statistical properties of objects in the real world or in the way we perceive them. It has not yet been proposed, however, how a plausible memory network could store reliably such representations; while attempts to model the storage of feature norms (experimentally obtained prototypes mimicking concept representations) with attractor networks have had success only using small sets of memories \cite{McRae_1997, McRae_1999, McRae_2006}. We propose here a way in which a purely Hebbian autoassociative memory could store and retrieve sets of correlated representations of any size, using a number of connections per neuron $C$ that increases proportionally with $p$. 

Interestingly enough, our learning rule is not quite appropriate for a one-shot learning process, since it requires to calculate statistical properties of the dataset - the popularity of neurons - {\em before} learning the patterns. In the case of semantic memory, concepts are acquired through a long time experience and through the repeated exposure to diverse versions of the input, allowing, if necessary, for a continuous updating of popularity estimates. Episodic memory, on the other hand, requires one-shot learning, leaving no time for a learning rule like ours to deal with the correlation between memories. Associative networks may have evolved in other directions to enable the on-line storage of episodes and events. Evidence has recently been obtained \cite{Leutgeb_2007} supporting the suggestion that the dentate gyrus acts as an orthogonalizing device in the heart of the medial temporal lobe episodic memory system \cite{Treves_1992}. The hippocampus could then function as an orthogonalized buffer, that helps neocortical networks acquire correlated memories through an off-line process. It has been proposed \cite{Marr_1971, Wilson_1994, Dayan_95} that it is during sleep that the hippocampus transfers to cortical areas the statistical biases of the input, in a process of \textit{consolidation}. While one-shot learning of a large dataset of orthogonal or randomly correlated patterns can be achieved through the `standard' rule of Eq. \ref{eq:oneshot}, the learning or stabilization of correlated memories in their final cortical destination may be consolidated by a learning process that reflects what in our model we have defined as the popularity of different neurons. Such consolidation may well accompany the spontaneous retrieval of representations stored in the hippocampus \cite{Squire_1991, McClelland_95}.

Our results show that correlated representations can be stored at a cost: memories lose homogeneity, some remaining robust and others becoming weak in an inverse relation to the information they convey. These side effects should be observed in any associative memory system that is understood to store correlated patterns directly, and absent if information is first equalized through pattern orthogonalization. 

Conversely, one may ask: are there benefits in representing correlated memories as they are, without recoding them into a more abstract, orthogonalized space? We have shown in a previous study \cite{NACO} that correlation plays a major role in driving a {\em latching} dynamics in a model of large cortical networks, in a process that could be a model of free association, and that might also underly the capacity for language \cite{Treves_2005}. Also, semantic priming has been shown to be guided by correlation \cite{Vigliocco_2004, McRae_1999}, selectively facilitating or inhibiting the retrieval of concepts, and potentially compensating for impaired episodic access \cite{Papik}. On the other hand, embodied theories of cognition suggest that far from creating a neural structure of its own, the semantic system evolved on the same neural substrates that already had a primary function (visual, tactile or motor processing, etc.), for which correlation in the representation, even if useful, would be an inevitable outcome of their history.

Some predictions of our theory could perhaps be tested experimentally. The most immediate result to test is the relationship between the distribution of patterns and their relative robustness. The distribution of neural activity of different memory representations is however not available, for obvious technical reasons. Imaging techniques do not offer the required resolution, and collecting adequate statistics from single unit recordings in animals appears prohibitive. Nevertheless, other measurable quantities could yield an estimate of relevant statistical properties of the distribution: priming effects, for example, are related to the correlations between memory items. A second way to test the theory could be to assess the retrieval of a memory by a partial cue, similarly to what has been proposed in \cite{Sartori_2004}, where the authors associate retrievability with a particular statistical measure: the \textit{semantic relevance} of the cue. A third possibility could be to measure the speed of retrieval, which can be related to Eqs. \ref{eq:mfinal} and \ref{eq:qfinal} and, again, to the specific cue that the network receives to trigger recall. In this last case, however, retrieval activity in the semantic system should be isolated from other processes, such as categorization, which could take place automatically, affecting the overall timing. Probing different systems other than semantic memory might also be a possibility, since our conclusions are general to any associative network with correlated memories. If a set of stimuli with controlled correlations were to be constructed (for example a set of pictures of caricature faces with exchangeable features), the memory of subjects trained with these stimuli could be tested for retrievability. The time-to-forget should then be related to the robustness, and inversely to the information content of each item, while with orthogonalized representations forgetting should be equalized.

\section{Methods}
\label{Methods}

\subsection{Sets of patterns used in simulations}

In the simulations shown in Fig. \ref{fig:scaling} a hierarchical algorithm was used to generate the patterns. The main idea is to produce, in the first place, a generation of random `parent' patterns which are not part of the dataset but are used to influence with different strength a second generation, $\{\xi^{\mu}\}$ (more details and a full analysis of the statistics of the resulting patterns can be found in \cite{NACO}). The reason to use this particular algorithm is that we needed a distribution of patterns with approximately the same correlation properties independently of $p$ and $N$. Following our studies in \cite{NACO}, this is the case with the above algorithm, as long as $p$ and $N$ are not too small and asymptotic statistics applies. 

For the simulations in Fig. \ref{fig:pfinal3} we needed higher levels of correlation than the ones that we could obtain with the algorithm described above, so as to illustrate the effects of large variability in the $S_f$ values of the patterns. On the other hand, we did not require in this case patterns with more than one value of $p$ and $N$. We then chose an algorithm that sets approximately an arbitrary popularity distribution over neurons. We chose
\begin{equation}
P(a_i)=\frac{1}{a}\exp \left( -\frac{a_i}{a} \right) \label{eq:dist},
\end{equation}
as the target distribution of popularity $F(x)$, with $\left<P(a_i)\right>\simeq a$. Since the total number of patterns is $p$, we defined the function
\begin{equation}
n_k=N P\left(k/p\right)
\end{equation}
expressing, when rounded to the closest integer, how many neurons should be active in $k$ patterns. For values of $n_k>0.5$, we assigned a target popularity $a_i=k/p$ to $round(n_k)$ arbitrary neurons. To construct each pattern $\mu$ we initially set all neurons in the pattern to be inactive. Then we picked neuron $i$ at random and set $\xi_i^{\mu}=1$ with probability $P_i$, until $aN$ neurons had been set to be active for each pattern. Finite size effects caused the actual distribution of popularity, shown in Fig. \ref{fig:pfinal3}a, to be slightly different from the target one in Eq. \ref{eq:dist}, specially for low values of popularity. Since this region of the distribution is the less interesting one (see Section \ref{general}), we did not modify the patterns further.

The feature norms analyzed in Fig. \ref{fig:mcrae} were downloaded from the \textit{Psychonomic Society Archive of Norms, Stimuli, and Data} web site, www.psychonomic.org/archive, with the consent of the authors. The norms list $p=541$ concepts relating several of $N=2526$ features to each one of them. To each concept we associated a $\mu$ index and to each feature a $i$ index. We set $\xi_i^{\mu}=1$ if feature $i$ was included in the description of pattern $\mu$ and $\xi_i^{\mu}=0$ otherwise. Since not all patterns are associated with the same number of features, the sparseness is not constant across patterns. The average sparseness is $a\simeq 0.006$ equivalent to $\sim 15$ features per concept. For each concept, $S_f$ is calculated as the average value of $a_i(1-a_i)$ among the features that comprise it.

\subsection{Testing the stability of memories}

The stability of a memory item should be tested irrespective of how accurate a cue it needs in order to be retrieved. For this reason, we used the full original pattern as a cue, which is a good approximation of its attractor. The initial state, thus, is set to coincide with the tested pattern. In each update step, a neuron $i$ is chosen at random and updated using the rule in Eq. \ref{eq:activation}, keeping track of $m$, whose initial value is close to $1$ by construction. Initially, $m$ varies rapidly, but it eventually converges to a stable value, either near $1$ or near $0$. A proof of this is the step like transition in the stable values of $m$, shown in Figure \ref{fig:pfinal3}b. The simulation stops when the variation of $m$ is smaller than a threshold, which we set small enough to give three digits accuracy in $m$.

\subsection{Storage capacity of more general distributions}
\label{general}
As we have shown in \textit{Results}, the important quantity to estimate in order to find the scaling of the storage capacity of a memory network with correlated patterns is the second term in the RHS of Eq. \ref{eq:qfinal}
\begin{equation}
\mathcal{T}_2=\frac{1}{a^2}\int_0^1 dx F(x)x(1-x)\phi \left(\frac{-U}{\sqrt{\alpha x q}}\right).
\end{equation}
The factor $\phi \left(-U/\sqrt{\alpha x q}\right)$ is $0$ when $x=0$ and reaches its maximum when $x=1$. On the other side, since we consider the sparse limit $a\ll 1$ the distribution $F(x)$ is concentrated toward small values of $x$. For these two reasons, the interesting part of any smooth distribution function $F(x)$ is the decay of its tail with increasing $x$. We study in this section two interesting cases: exponential and power-law distributions. Keeping in mind that the exact behavior of $F(x)$ for small values of $x$ is less relevant, these results can be generalized to any distribution function with such tails. 
\subsubsection{Exponential distribution}
\label{exp}
The exponential distribution
\begin{equation}
F(x)=\frac{\exp(-x/a)}{a}
\end{equation}
is normalized to $1$ and has mean equal to $a$ -- apart from a small correction of order $\exp(-1/a)$, which we neglect for simplicity. Its variance is about $a^2$, with a correction of the same order. Finally, $S_f\simeq a(1-2a)$. The critical second term in the RHS of Eq. \ref{eq:qfinal} is
\begin{equation}
\mathcal{T}_2=\frac{1}{a^2}\int_0^1dx F(x) x(1-x)\int_{-\infty }^{\sqrt{y/x}}Dz=\frac{1}{a^2}\int_{\sqrt{y}}^{\infty }Dz\int_{y/z^2}^1 dx \frac{\exp(-x/a)}{a} x(1-x)
\end{equation}
where we have inverted the integration order. $Dz$ is the gaussian differential defined in Eq. \ref{eq:gauss} and $y=U^2a/(\alpha S_f)$. The inner integral in the right-most side of the equation confirms that the value of $F(x)$ for small $x$ is less relevant than its decay for large $x$. The RHS is now integrable, resulting in
\begin{equation}
\mathcal{T}_2=\frac{1}{a^2}\int_{\sqrt{y}}^{\infty }dz \frac{1}{\sqrt{2\pi}}\exp\left(-\frac{z^2}{2}-\frac{y}{a z^2}\right)\left[S_F+\frac{y}{z^2}\left(1-\frac{y}{z^2}-2a\right)\right].
\end{equation}
This expression can be integrated a second time, but its analytical expression is too complicated to include here. It is enough to mention that the largest contribution is proportional to $\exp\left(-\sqrt{2y/a}\right)$

\begin{equation}
\mathcal{T}_2\simeq \frac{1}{2a^2}\exp\left(-\sqrt{\frac{2y}{a}}\right)\left(S_F+\sqrt{\frac{a y}{2}}-\frac{a}{2}\sqrt{\frac{a y}{2}}+\frac{a y}{2}-2a \sqrt{\frac{a y}{2}}\right) .
\end{equation}
Assuming $2y/a\sim 1$ modulo some logarithmic correction (that we consider inside the exponential and neglect elsewhere) this results in
\begin{equation}
\mathcal{T}_2\simeq \exp\left(-\sqrt{\frac{2y}{a}}\right)\frac{3}{4a^2}S_F.\label{eq:t2}
\end{equation}
Since only $y$ depends on $\alpha_c$ it is easy to see from this equation that indeed $2y/a\sim 1$ modulo logarithmic corrections, making the previous assumption self-consistent. The storage capacity can be obtained by making the RHS of Eq. \ref{eq:t2}, as in the previous section, equal to $S_f/a$, 
\begin{equation}
\alpha_c\simeq\frac{2U^2}{S_f\left[ \ln\left(\frac{3S_F}{4aS_f}\right)\right]^2}\propto \frac{1}{S_f\left[\ln\left(\frac{S_F}{aS_f}\right) \right]^2}.
\end{equation}
Note that the square on the logarithmic factor makes this storage capacity lower than the one found for $F(x)$ distributions of very low variance. Again, the correction is valid as long as the logarithm is large, in other words $\ln\left(S_F/aS_f\right)>1$. If this condition is not met, the storage capacity scales like $1/S_f$.

\subsubsection{Power law distribution}

We define the power law distribution

\begin{equation}
F(x)=\left\{\begin{array}{ll}
0  & \mbox{if $x<d$} \\
c x^{-\gamma}  & \mbox{if $x>d$}
\end{array}
\right.
\end{equation}
with $\gamma>2$ and $d$ a small cutoff value that prevents the integral of $F(x)$ from diverging. The conditions for normalization and mean are
\begin{equation}
1=c \left(\frac{d^{1-\gamma}-1}{\gamma-1} \right)
\end{equation}
\begin{equation}
a=c \left(\frac{d^{2-\gamma}-1}{\gamma-2} \right).
\end{equation}
There is no simple analytical expression for $c$, $d$ or $S_F$ in terms of $a$ and $\gamma$.

We want to compute
\begin{equation}
\mathcal{T}_2=\frac{1}{a^2}\int_d^1dx \ c\ x^{-\gamma}x(1-x) \phi\left(-\sqrt{\frac{y}{x}}\right)
\end{equation}
where, again, $y=U^2a/(\alpha S_f)$. $\mathcal{T}_2$ is integrable, resulting in

\begin{eqnarray}
\mathcal{T}_2 & = & \frac{c}{a^2}\phi\left[-\sqrt{y}\right]\left(\frac{1}{\gamma-3}-\frac{1}{\gamma-2}\right)+\frac{c}{a^2}\phi\left[-\sqrt{\frac{y}{d}}\right]d^{2-\gamma}\left(\frac{d}{\gamma-3}-\frac{1}{\gamma-2}\right) -\nonumber \\
& - & \frac{c}{a^2(\gamma-3 )}\left(\frac{1}{2\sqrt{\pi}}\left(\frac{y}{2}\right)^{3-\gamma}\left\{\Gamma\left[-\frac{5}{2}+\gamma,\frac{y}{2}\right]-\Gamma\left[-\frac{5}{2}+\gamma,\frac{y}{2d}\right]\right\}\right) +\nonumber \\
& + & \frac{c}{a^2(\gamma-2 )}\left(\frac{1}{2\sqrt{\pi}}\left(\frac{y}{2}\right)^{3-\gamma}\left\{\Gamma\left[-\frac{3}{2}+\gamma,\frac{y}{2}\right]-\Gamma\left[-\frac{3}{2}+\gamma,\frac{y}{2d}\right]\right\}\right)
\end{eqnarray}
where $\Gamma[,]$ is the incomplete gamma function. The following series expansions are useful
\begin{eqnarray}
\phi[-\sqrt{y}]& = &\frac{\exp (-\frac{y}{2})}{\sqrt{2 \pi y}} \left\{ 1+\sum_{k=1}^{\infty }\left[\prod_{j=1}^k(2j-1)\right](-y)^{-k}\right\} \\
\frac{1}{2\sqrt{\pi}}\left(\frac{y}{2}\right)^{n-\gamma} \Gamma \left[-n+\frac{1}{2}+\gamma ,\frac{y}{2} \right] & = & \frac{\exp (-\frac{y}{2})}{\sqrt{2 \pi y}} \left\{1+\sum_{k=1}^{\infty }\left[\prod_{j=1}^k(2j-1+2(n-\gamma))\right](-y)^{-k}\right\}.\nonumber 
\end{eqnarray}
$\mathcal{T}_2$ is different from $0$ only to order $y^{-2}$ inside the curly brackets. At this order of approximation
\begin{equation}
\mathcal{T}_2\simeq \frac{4 c \exp \left(-y/2 \right)}{a^2\sqrt{2\pi y^5}}
\end{equation}
neglecting a similar term including the factor $\sqrt{d^5}\exp \left(-\frac{y}{2d} \right)$. As previously, the storage capacity can be estimated as
\begin{equation}
\alpha_c\propto \frac{a }{S_f\ln\left(\frac{a^{\gamma-2}}{S_f}\right)}
\end{equation}
where we have used $c\propto a^{\gamma-1}$. If the logarithm is of order $1$ or smaller the storage capacity scales simply like $a/S_f$.

\section{Acknowledgments}
This research was supported by Human Frontier Science Program Grant RGP0047/2004-C. 
\section{References}
\bibliographystyle{apalike}
\bibliography{correlations}

\end{document}